\documentclass[pra,showkeys,showpacs,groupedaddress,twocolumn]{revtex4}
\usepackage{amssymb}
\usepackage{graphicx}
\usepackage{dcolumn}
\usepackage{bm}
\usepackage{amsmath}
\usepackage{subfigure}
\usepackage{float}
\usepackage{color}

\begin{document}

\title{Dipolar Rydberg-atom gas prepared by adiabatic passage through an avoided crossing}

\author{Limei Wang}
\author{Hao Zhang}
\author{Linjie Zhang}
\author{Changyong Li}
\author{Yonggang Yang}
\author{Jianming Zhao}
\thanks{Corresponding author: zhaojm@sxu.edu.cn}
\author{Georg Raithel$^{1}$}
\author{Suotang Jia}
\affiliation{ State Key Laboratory of Quantum Optics and Quantum Optics Devices, Institute of Laser spectroscopy, Shanxi University, Taiyuan 030006, P. R. China}
\affiliation{$^{1}$ Department of Physics, University of Michigan, Ann Arbor, Michigan 48109-1120, USA}

\begin{abstract}
The passage of cold cesium 49S$_{1/2}$ Rydberg atoms through an electric-field-induced multi-level avoided crossing with nearby hydrogen-like Rydberg levels is employed to prepare a cold, dipolar Rydberg atom gas. When the electric field is ramped through the avoided crossing on time scales on the order of 100~ns or slower, the 49S$_{1/2}$ population adiabatically transitions into
high-\emph{l} Rydberg Stark states. The adiabatic state transformation results in a cold gas of Rydberg atoms with large electric dipole moments. After a waiting time of about $1~\mu$s and at sufficient atom density, the adiabatically transformed highly dipolar atoms become undetectable, enabling us to discern adiabatic from diabatic passage behavior through the avoided crossing. We attribute the state-selectivity to $m$-mixing collisions between the dipolar atoms. The data interpretation is supported by numerical simulations of the passage dynamics and of binary $m$-mixing collisions.
\end{abstract}
\keywords{adiabatic passage, avoided crossing, Rydberg-atom collisions, dipolar atomic gas}
\pacs{32.80.Xx, 32.80.Ee, 34.90.+q}

\maketitle

\section{Introduction}

Atoms in highly excited Rydberg states (principal quantum number $n$) have large radii and electric-dipole transition matrix elements ($\sim n^{2}$), large polarizabilities ($\sim n^{7}$) and strong van-der-Waals interactions ($\sim n^{11}$)~\cite{cambridge}.These properties have led to a variety of interesting investigations and applications, including quantum information and logic gates~\cite{galindo, garcia-ripoll, jaksch, fleischhauer, isenhower}, single-photon sources~\cite{dudin} enabled by the Rydberg excitation blockade effect~\cite{fleischhauer,comparat,tong,vogt,vogtviteaut}, and many-body physics with strong long-range interactions~\cite{and,muller,olmos,rev,bariani}. The large polarizability makes Rydberg atoms sensitive to external fields, giving rise to applications in field measurement~\cite{Mohapatra,Osterwalder}, quantum control~\cite{nipper} and studies involving collisions~\cite{cambridge} and novel molecules~\cite{marcassa}.

\begin{figure}[b]
\vspace{-1em}
\centering
\includegraphics[width=0.5\textwidth]{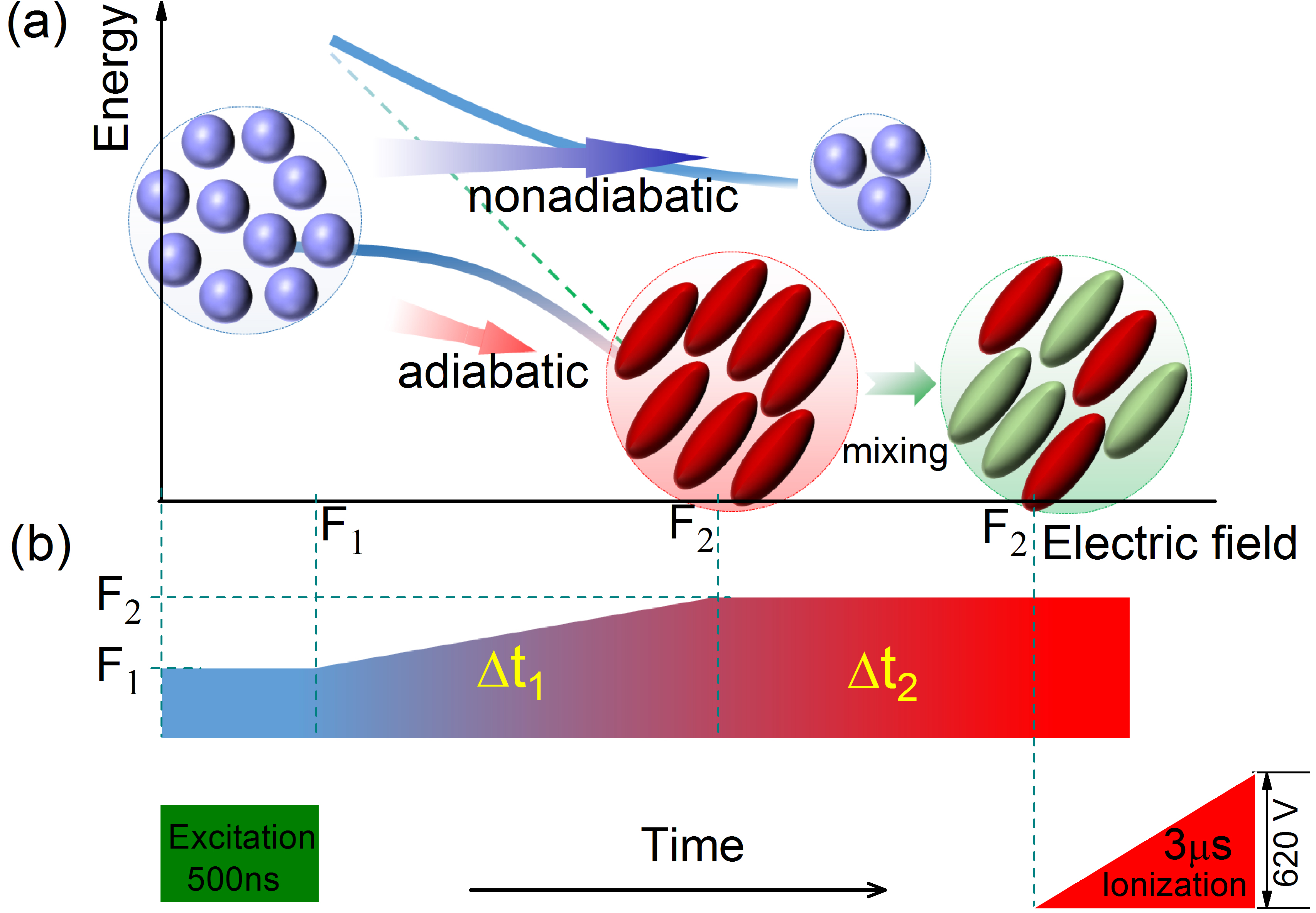}
%\vspace{-1em}
\caption{(Color online) Passage of Rydberg atoms
through an electric-field-induced avoided crossing and the resultant state-mixing properties.
(a) Rydberg atoms in a state close to
49S$_{1/2}$ (circles) are prepared at an electric field $F_{1}$ = 3.14~V/cm. The field is then linearly ramped
to $F_{2}$ = 3.99~V/cm with a rise time $\Delta t_1$ across a selected avoided crossing. The Rydberg atoms undergo adiabatic/diabatic passage through the avoided crossing. Atoms passing adiabatically are transformed into an $m=1/2$ high-$\langle \emph{l} \rangle$ elongated Stark state (ovals). During a hold time $\Delta t_2$, they undergo efficient \emph{m}-mixing into Stark states with high $\vert m \vert$ (ovals of different shades).
(b) Timing diagram. After the hold time $\Delta t_2$, a ramp ionization field (except specified otherwise) is applied that only ionizes the atoms in low-$\vert m \vert$ levels but not the ones in high-$\vert m \vert$ levels.}
\end{figure}

We investigate how a controlled passage of a dense cloud of Rydberg atoms through an avoided crossing alters the collision-induced dynamics of the sample. As a result of adiabatic state transformation, Rydberg atoms passing adiabatically acquire large permanent electric dipole moments, which lead to enhanced dipole-dipole interactions. The accelerated  state mixing is probed via time-delayed state-selective field ionization~\cite{cambridge,HJD,AGuertler}.
Previously, the (single-atom) dynamics of lithium Rydberg atoms passing through an avoided crossing has been studied via a measurable difference in the ionization electric fields of the atoms exhibiting diabatic and adiabatic passage~\cite{rubbmark}. Rydberg \emph{l}-state redistribution has been controlled by application of a large-amplitude rectangular electric field pulse~\cite{tannian}. In collision studies, \emph{l}-changing interactions of Na $n$D Rydberg atoms with slow Na$^{+}$ ions have been investigated using field-ionization templates for \emph{l} =3, 4, and 5~\cite{sun}. The effect of \emph{l} and \emph{m}-mixing by weak, homogeneous dc electric fields and static ions has been predicted to lengthen the lifetimes of Rydberg states \cite{FMerktandRNZare}. The mixing-induced long lifetimes of high-$n$ Rydberg states play a central role in zero electron kinetic energy (``Zeke'') spectroscopy~\cite{FMerkt,kmullerd}. In cold Rydberg-atom gases, plasma formation in a cloud of initially low-\emph{l} Rydberg atoms and subsequent recombination processes can generate high-\emph{l} Rydberg atoms~\cite{robinson}. Long-lived high-\emph{l} Rydberg states have been created by \emph{l}- and \emph{m}-mixing collisions in rubidium~\cite{dutta} and by state transfer induced with weak electric fields in cesium~\cite{zhang}.

Here, we employ the adiabatic/diabatic passage of cesium Rydberg atoms in a well-defined initial state, prepared within an applied electric field, through a selected multi-level avoided crossing. In alkali-metal atoms,
low-\emph{l} (\emph{l}$<$2) Rydberg states typically have low electric-dipole moments while the outermost linear Stark levels have large ones, resulting in sequences of avoided crossings between low-\emph{l} states and linear, high-$\langle \emph{l} \rangle$ Stark states~\cite{zimmerman}. Time-dependent studies of avoided crossings\cite{hda} are of general relevance, since avoided crossings are a universal phenomenon in atoms and molecules. Here, we are interested in the dynamics of cold, relatively dense Rydberg atom samples after transformation of the Rydberg-atom wavefunction in the avoided crossing. In our case, adiabatic transformation induces large permanent electric dipole moments, which have a profound effect on the subsequent collision-induced dynamics of the samples. We vary the speed of the electric-field ramp that drives the atoms through the avoided crossing as well as the atom interaction time after completion of the ramp. The final atom sample is analyzed via state-selective field-ionization. We interpret our results based on simulations of the passage behavior and the collision properties of the sample after the passage.

\section{Setup}

We trap cesium atoms in a standard magneto-optical trap (MOT) with temperature of $\sim$100 $\mu$K and peak density
of $\sim 10^{10}$~cm$^{-3}$, and use stepwise two-photon excitation to prepare Rydberg states. The trapping laser is turned off during Rydberg-atom excitation, manipulation and detection, whereas the repumping laser is left on to avoid optical pumping into 6S$_{1/2}$(F=3) by off-resonant transitions. The lower-transition laser resonantly drives the 6S$_{1/2}$ (F=4) $\to$  6P$_{3/2}$ (F$'$=5) transition and has a power of 660~$\mu$W and a Gaussian beam waist of $w_0 \sim 1.2$~mm.
The upper-transition laser drives a transition from 6P$_{3/2}$ (F$'$=5) into a Rydberg state and has a waist diameter of $w_0 \sim 50~\mu$m, yielding a cylindrical excitation volume with a length of $\sim$ 800$~\mu$m and a diameter of $\sim$ 50$~\mu$m.
The excitation-pulse duration is 500~ns. In order to reproducibly excite selected, well-defined Rydberg levels under presence of an applied electric field, where the density of states is large, we use a wavelength meter with an uncertainty of 30~MHz. The Rydberg atoms are ionized using state-selective electric-field ionization~\cite{cambridge}. The released ions are detected
with a calibrated micro-channel plate (MCP). Since the MCP ion detection efficiency is $\sim 35\%$, the actual Rydberg atom numbers and densities are about a factor of three higher than the numbers and densities of detected Rydberg atoms.

Electric fields are applied using a pair of
parallel, non-magnetic grids centered at the MOT location (spacing 15~mm). The electric fields are generated using an arbitrary-waveform generator (voltage precision $\pm$1~mV). The field is calibrated and zeroed via Stark spectroscopy of 60D$_{3/2}$ and 60D$_{5/2}$. The experiment and the time sequence are sketched in Fig.~1.

\section{Spectroscopy}

\begin{figure}[htb]
\vspace{-2em}
\centering
\includegraphics[width=0.4\textwidth]{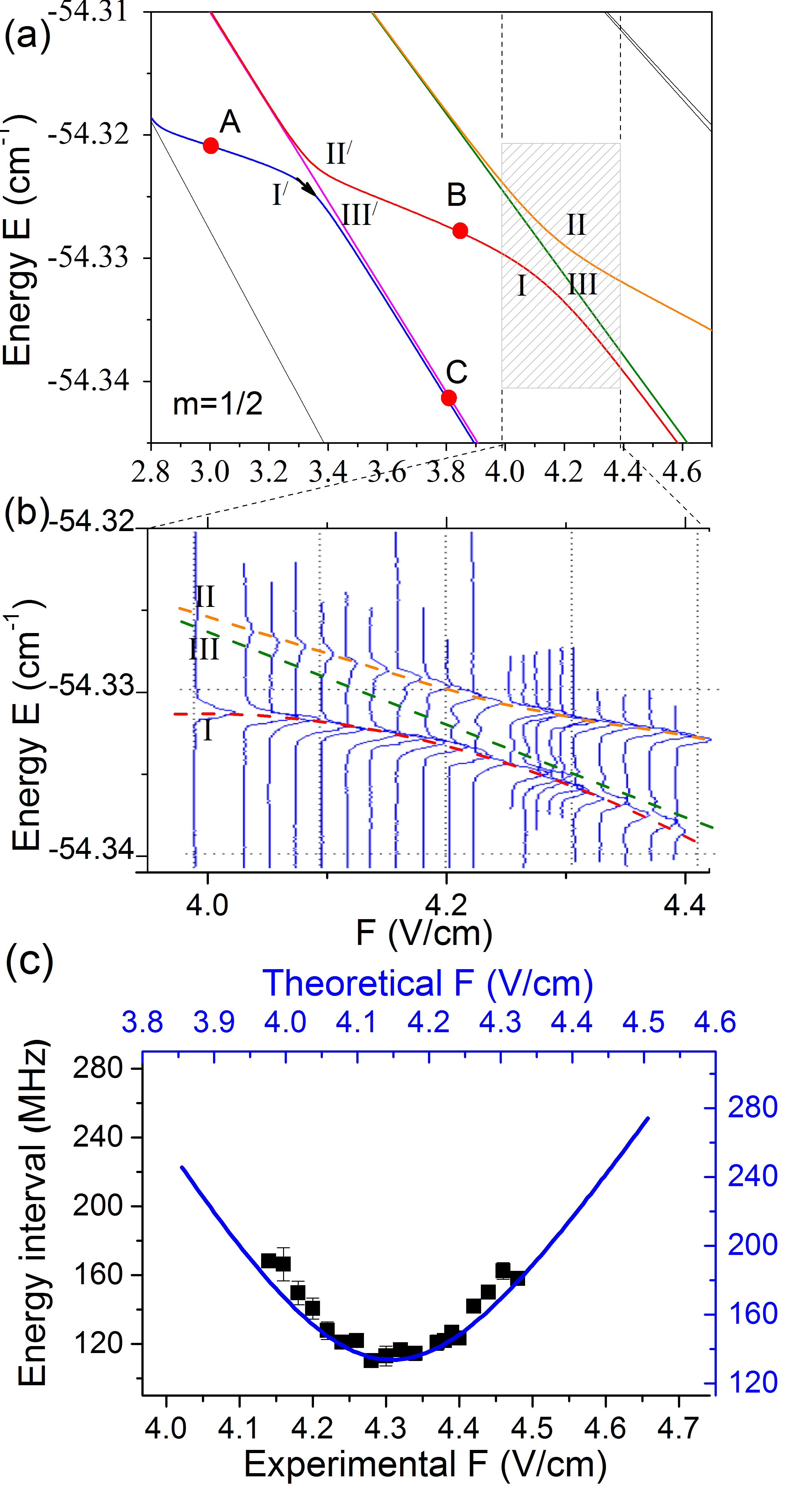}
%\vspace{-3em}
\caption{(Color online) Theoretical Stark map (a), measured spectra for a set of electric fields (b) and according energy gaps between states $|I \rangle$ and $|II \rangle$ (c) of avoided crossings between the 49S$_{1/2}$ state and the $n=45$ manifold of states in cesium. Each avoided crossing involves three adiabatic states, labeled $|I \rangle$, $|II \rangle$, $|III \rangle$ etc.. The points A and B in the Stark map identify Stark states that have primarily 49S$_{1/2}$ character and are located between two adjacent avoided crossings. The point C identifies the high-$\langle \emph{l} \rangle$ Stark state that becomes populated via adiabatic passage in our experiments. The line in (c) shows the result of the calculation (top and right axes).}
\end{figure}

In preparation for our time-dependent studies, it is important to locate a suitable avoided crossing, to verify that the field inhomogeneity is low enough that the crossing can be well resolved, and to measure its gap size $\Delta E$. For our work we have calculated~\cite{zimmerman} and measured cesium Stark spectra in the vicinity of the $n=45$ hydrogen-like manifold of states; respective results are shown in Figs.~2~(a) and~(b).
The Stark map in (a) includes two three-level avoided crossings with adiabatic states labeled $|I \rangle$,
$|II \rangle$, $|III \rangle$ and $|I' \rangle$, $|II' \rangle$, $|III' \rangle$; the respective energies are E$_{I}$, E$_{II}$, E$_{III}$ and
E$_{I^{'}}$, E$_{II^{'}}$, E$_{III^{'}}$. The measured spectrum in Fig.~2~(b) shows atom counts as a function of wavenumber for a selection of electric-field values, corresponding to the shaded field and energy range in Fig.~2~(a). Both measurement and calculation demonstrate that the avoided crossings involve three atomic levels, whereby the line strength of the middle adiabatic level $|III \rangle$ is much smaller than that of the $|I \rangle$ and $|II \rangle$-levels.

The avoided crossings are characterized by a center electric field, $F_{X}$, and the energy gap at $F_{X}$.
For each avoided crossing, $F_{X}$ is given by the electric-field value at which the energy gap, $\Delta E $, between the adiabatic levels $|I \rangle$ and $|II \rangle$ exhibits a (local) minimum as a function of electric field $F$, {\sl{i.e.}}
\begin{equation}
\Delta E = \min_{F} (E_{II}(F)-E_{I}(F)) \quad
\end{equation}
the gap minimum is realized when $F=F_X$.
As an example, in Fig.~2~(c) we show the experimental energy-level difference  $\Delta E $ (symbols) and calculation (solid line) between states $|I \rangle$ and $|II \rangle$ as a function of electric field for the anti-crossing in Fig.~2~(a).
The measured center field is $F_X=4.3$~V/cm and the gap minimum $\Delta E (F_X) = 110$~MHz.
The measured energy gap agrees with the calculation within about $20\%$, and the field within $4\%$.
The intermediate level $E_{III}$ exhibits no significant curvature, leading us to infer that its coupling to the other levels is small and that its effect on $\Delta E$ is minor. This is confirmed below in Sec.~\ref{sec:lztheory}.

\section{Passage behavior in ramped electric fields}

In our time-dependent studies we employ the avoided crossing that involves the adiabatic states $|I' \rangle$, $|II' \rangle$ and $|III' \rangle$ defined in Fig.~2~(a); this avoided crossing has a gap size, $\Delta E = 60$~MHz. We initially prepare Rydberg atoms in adiabatic state $|I' \rangle$ at point A in Fig.~2~(a), at an electric field of 3.14~V/cm. As the electric field is linearly ramped through the avoided crossing centered at $F_{X} \approx 3.4$~V/cm to a final field of 3.99~V/cm, the atoms undergo diabatic or adiabatic passage into adiabatic states $|II' \rangle$  and $|I' \rangle$ at points B and C, respectively. The process strongly depends on the ramp time, $\Delta t_{1}$, of the electric field. When the electric field varies fast, the atoms preferentially pass diabatically into level $|II' \rangle$ at point B, while for
slow ramps the population mostly passes adiabatically into level $|I' \rangle$  at point C [black arrow shown in Fig.~2~(a)].
The 49S$_{1/2}$-character of the adiabatic quantum states at points A and B in Fig.~2~(a) exceeds $90\%$, while the adiabatic state at point C primarily contains hydrogenic states with $l \gg 0$. We therefore refer to the state at C as high-$\langle \emph{l} \rangle$ (note, however, that the quantum number \emph{l} is not conserved in non-zero fields). Ignoring level $|III' \rangle$ and noting that the anti-crossings are sufficiently well separated that interactions between them~\cite{hda} are insignificant, the passage approximately follows Landau-Zener dynamics~(see Ref.~\cite{rubbmark} and Sec.~\ref{sec:lztheory}).

 \begin{figure}[htb]
%\vspace{-3em}
\centering
\includegraphics[width=0.4\textwidth]{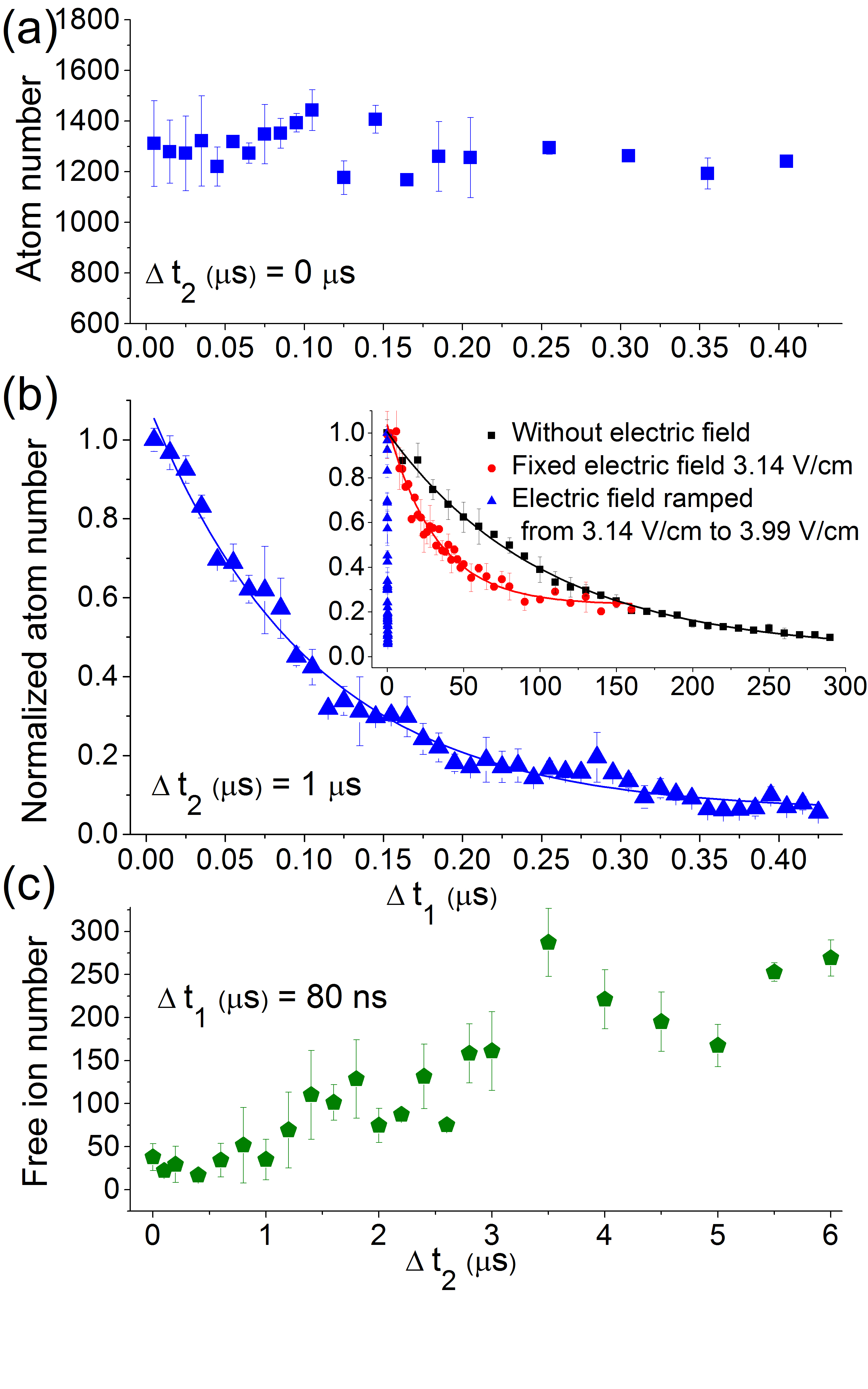}
%\vspace{-3em}
\caption{(Color online) (a) Measured atom number as a function of ramp time $\Delta t _{1}$. The atoms are initially prepared in point A in Fig.~2, and the electric field is ramped  from 3.14 to 3.99~V/cm (through one of the avoided crossings in Fig.~2). The pulsed ionization electric field is set high enough that 49S$_{1/2}$-like atoms (point A in Fig.~2) are detected, but it is too low to detect high-$\vert m \vert$ atoms ($\vert m \vert \gtrsim 2.5$). The Rydberg-atom density is $N _{0}=3 \times 10^{9}~$cm$^{-3}$ and the holding time $\Delta t _{2}$ = 0~$\mu$s. (b) Normalized atom number for a holding time of $\Delta t _{2}$ = 1~$\mu$s (other conditions same as in (a)). The solid line shows an exponential fit to the data.
The inset in (b) shows the decay behavior of pure 49S$_{1/2}$-atoms at zero electric field (black squares) and of 49S$_{1/2}$-like atoms at a fixed field of 3.14~V/cm (point A in Fig.~2; red circles). (c) Detected number of Penning-ionized atoms as a function of hold time $\Delta t_{2}$ for  fixed $\Delta t_{1}$ = 80~ns and same density as in (a) and (b).}
\end{figure}

Atoms adiabatically transferred into point C in Fig.~2~(a) still have the same $m$ as the initially prepared 49S$_{1/2}$-like state at point A, and are energetically very close to the 49S$_{1/2}$-like state at point B (which is populated via diabatic passage). Therefore, atoms in points B and C cannot be directly distinguished by state-selective field ionization (in contrast to the case studied in~Ref.~\cite{rubbmark}, where such a distinction was possible). This is shown in Fig.~3~(a), where we plot the counted atom number as a function of ramp time $\Delta t_{1}$ for a field ionization voltage that is set just high enough to ionize the initially populated 49S$_{1/2}$-like atoms points A and B in Fig.~2 (a)). Based on our calculation of the passage probabilities in Sec.~\ref{sec:lztheory}, in Fig.~3~(a) the adiabatic-passage probability changes from near zero to near 100$\%$ over the
investigated range of $\Delta t_1$, with no resultant discernable trend in the atom counts. We conclude that field ionization immediately after the ramp through the avoided crossing is not suitable to state-selectively detect atoms in points B and C.

\section{Collision-induced dynamics caused by adiabatic state transformation}
\label{sec:data}

Adiabatic passage transforms the atoms from weak into strong dipolar character. Atoms in points A and B have electric dipole moments from about 130~e$a_{0}$ to 160~e$a_{0}$, while atoms in point C have dipole moments of about 900~e$a_{0}$. This is interesting in several respects. Direct optical excitation of such a sample would be difficult to accomplish because of optical selection rules (atoms in point C have only about $1\%$ 49S$_{1/2}$-character) and because the Rydberg excitation blockade~\cite{comparat} suppresses optical or microwave excitation of high-density Rydberg-atom samples at point C (due to interaction-induced electric-dipole energy shifts). Hence, atom samples prepared by adiabatic passage from A to C enable us to study collision-induced dynamics in dipolar Rydberg atom gases under Rydberg-atom density conditions that would likely not be attainable using alternate means.

Enhanced dipolar interactions change the sample dynamics after passaging through the avoided crossing. In Fig.~3~(b) we hold the field for a time $\Delta t_{2}=1~\mu$s at its final value and show the normalized number of detected atoms as a function of the ramp time $\Delta t_{1}$ at a Rydberg-atom density of $N_0= 3\times 10^{9}$~cm$^{-3}$ (blue triangles). As in Fig.~3~(a), we
use a pulsed ionization field with voltage just high enough to ionize the initially populated 49S$_{1/2}$-like atoms. Over the investigated range of $\Delta t_{1}$ the Landau-Zener adiabatic-passage probability changes from near zero to near 100$\%$ (see Sec.~\ref{sec:lztheory} below). In contrast to
Fig.~3~(a), where the detected atom fraction is fairly constant at 100$\%$, in Fig.~3~(b) the fraction of detected atoms drops from 100$\%$ to near zero, as the adiabatic transition probability increases. It is concluded that the waiting time $\Delta t_{2}=1~\mu$s enables state-selective detection of adiabatic and diabatic passage of atoms through the avoided crossing. The adiabatically transformed atoms become undetectable over a wait time of $\Delta t_2 \sim 1~\mu$s. The signal decay observed in Fig.~3~(b)
is fitted quite well by a decaying exponential with a decay time constant of 103~ns $\pm$ 5~ns.
To explain the observed behavior, we first note that the atoms in points $B$ and $C$ in Fig.~2 (a) are nearest-neighbor quantum states of the same $m=1/2$ Stark map. Such states can usually not be distinguished by using state-selective field ionization such as ours; therefore the result in Fig.~3~(a) is not surprising. However, if a sufficiently long hold time $\Delta t_{2}$ is introduced, the atoms in highly dipolar, elongated Stark states (point C in Fig.~2 (a)) may selectively undergo $m$-mixing collisions into energetically nearby Stark states with higher $\vert m \vert$. As $\vert m \vert$ exceeds a value of about 2.5, atoms tend to ionize diabatically because the gaps of the avoided crossings in the Stark maps trend towards zero~\cite{Kleppner1980}. The ionization electric fields then increase by factors between 1.8 and 4 (for a calculation of hydrogenic ionization rates see~\cite{Damburg}). Hence, state-selective, efficient $m$-mixing and a corresponding increase in ionization field are the most likely reasons of why the adiabatically transformed atoms become undetectable. The results in Fig.~3 (b) for short $\Delta t_{1}$, where the passage is diabatic, further indicate that atoms in $49S_{1/2}$-like states (point B in Fig.~2 (a)) do not exhibit sufficient state mixing to alter the field ionization threshold. A state-mixing model is investigated in detail in Sec.~\ref{sec:mmixing} below.

To support the above interpretation of the data, in the inset of Fig.~3~(b) we excite 49S$_{1/2}$-like atoms at a fixed electric field of 3.14~V/cm (point A in Fig.~2 (a)) and Rydberg atom density $N_{0}$, and we show the normalized detected atom number as a function of time $\Delta t_1$ over a very long range up to 300~$\mu$s (red circles). For comparison, we also show the decay of 49S$_{1/2}$ atoms in zero field (black squares). The decays of the pure 49S$_{1/2}$-atoms at zero field and the 49S$_{1/2}$-like atoms at point A in Fig.~2~(a) both occur on time scales that exceed the time scale found in the main panel of Fig.~3~(b) by about two orders of magnitude (to visualize this stark contrast, the data from the main panel in Fig.~3~(b) are also shown in the inset). Hence, the adiabatic passage through the avoided crossing and the dynamics that follows during the subsequent $\sim 1~\mu$s greatly accelerate the signal decay.

Less important but noteworthy are the following observations. The decay of the pure 49S$_{1/2}$-atoms is exponential (see fitted curve in the inset of Fig.~3~(b)) and has a fitted decay time of $103~\mu$s $\pm$ $4~\mu$s at zero field. This value is between the level lifetimes calculated for near-zero (0.1~K) radiation temperature, where it is $120~\mu$s, and for 300~K radiation temperature, where it is $60~\mu$s. The chamber is at room temperature. The calculated lifetime at 300~K includes black-body-driven microwave and THz transitions into other Rydberg levels. Such transitions are not registered as decay events in the long-time data in the inset of Fig.~3~(b), which are obtained with a higher ionization electric field than the other data in Fig.~3.
Hence, the experimentally observed decay time is indeed expected to be between the calculated level lifetimes at 0.1~K and at 300~K.
Further,  49S$_{1/2}$ atoms in zero field exhibit only weak van-der-Waals forces and vanishing $m$-mixing; hence any collision-induced effects are expected to be small even at long times. The observed exponential decay corroborates this expectation. The calculated level decay time for the 49S$_{1/2}$-like atoms at point A in Fig.~2~(a) also is about $60~\mu$s. In the experiment, the decay of the 49S$_{1/2}$-like atoms at point A is non-exponential, with an accelerated initial decay time of $32~\mu$s
and a longer-lived fraction of atoms remaining at times past $\sim150~\mu$s. This indicates that initially the decay of the 49S$_{1/2}$-like atoms at point A is enhanced by collisions.  The 49S$_{1/2}$-like atoms at point A have an electric-dipole moment of 130~e$a_{0}$, which results in significant interatomic forces. Over the long time scale covered in the inset in Fig.~3~(b), these forces are likely to cause $n$-mixing collisions and Penning ionization~\cite{HNakamura,AReinhard}. These collisions may cause the initial accelerated decay of the 49S$_{1/2}$-like atoms at point A, in comparison with pure 49S$_{1/2}$ atoms in zero field. A fraction of the product Rydberg atoms are in long-lived high-$\vert m \vert$ states, which survive for much longer times~\cite{dutta}.

Finally, to support our interpretation further, in Fig.~3~(c) we show the number of free ions generated by Penning ionization as a function of $\Delta t_2$ for a fixed ramp time $\Delta t_1 = 80~$ns, which is sufficiently long for the passage to be $>90\%$ adiabatic (see Sec.~\ref{sec:lztheory}). While for the wait time $\Delta t_2 = 1~\mu$s used in Fig.~3~(b) the ion fraction is only about 2~$\%$, it rises to about 20~$\%$ at $\Delta t_2 = 6~\mu$s. The fractions of atoms participating in Penning-ionizing collisions are twice the ion fractions. Thermal ionization due to black-body radiation is negligible over the time range in Fig.~3~(c).
The fact that Penning-ionizing collisions are seen in Fig.~3~(c) makes it very plausible that $m$-mixing collisions, which are near-elastic and have larger rates than Penning ionization, are highly probable for a wait time $\Delta t_2 = 1~\mu$s.

We have established that two conditions are necessary for the 49S$_{1/2}$-like atoms to become undetectable. Firstly, the passage through the avoided crossing needs to be adiabatic. From Fig.~3~(b) and Sec.~\ref{sec:lztheory} it follows
that adiabaticity requires $\Delta t_1 \gtrsim 100~$ns. Secondly, our experiments show that a wait time $\Delta t_2 \gtrsim 1~\mu$s is needed for the 49S$_{1/2}$-like atoms to undergo $m$-mixing and to acquire an ionization electric field that is larger than the applied ionization field. The requirement on the wait time $\Delta t_2$ is supported by the $m$-mixing calculation presented in Sec.~\ref{sec:mmixing}.

\section{Density dependence of the collision-induced dynamics}

In Sec.~\ref{sec:data} we have attributed the signal decay observed in Fig.~3~(b) to adiabatic passage and subsequent \emph{m}-mixing of atoms in point C of Fig.~2 (a) during the hold time of $\Delta t_{2}$ = 1~$\mu$s. In order to test whether \emph{m}-mixing collisions are essential for atoms in point C to become undetectable, we have taken data equivalent to those in Fig.~3~(b) and (c) at lower densities. The results shown in Fig.~4~(a) clearly demonstrate that the
loss in signal becomes less significant with decreasing atom density. Also, the time scale over which the signal loss develops increases with decreasing density. Both observations are consistent with collisions playing a central role for atoms in point C of Fig.~2 (a) to become undetectable.

\begin{figure}[htb]
%\vspace{-3em}
\centering
\includegraphics[width=0.4\textwidth]{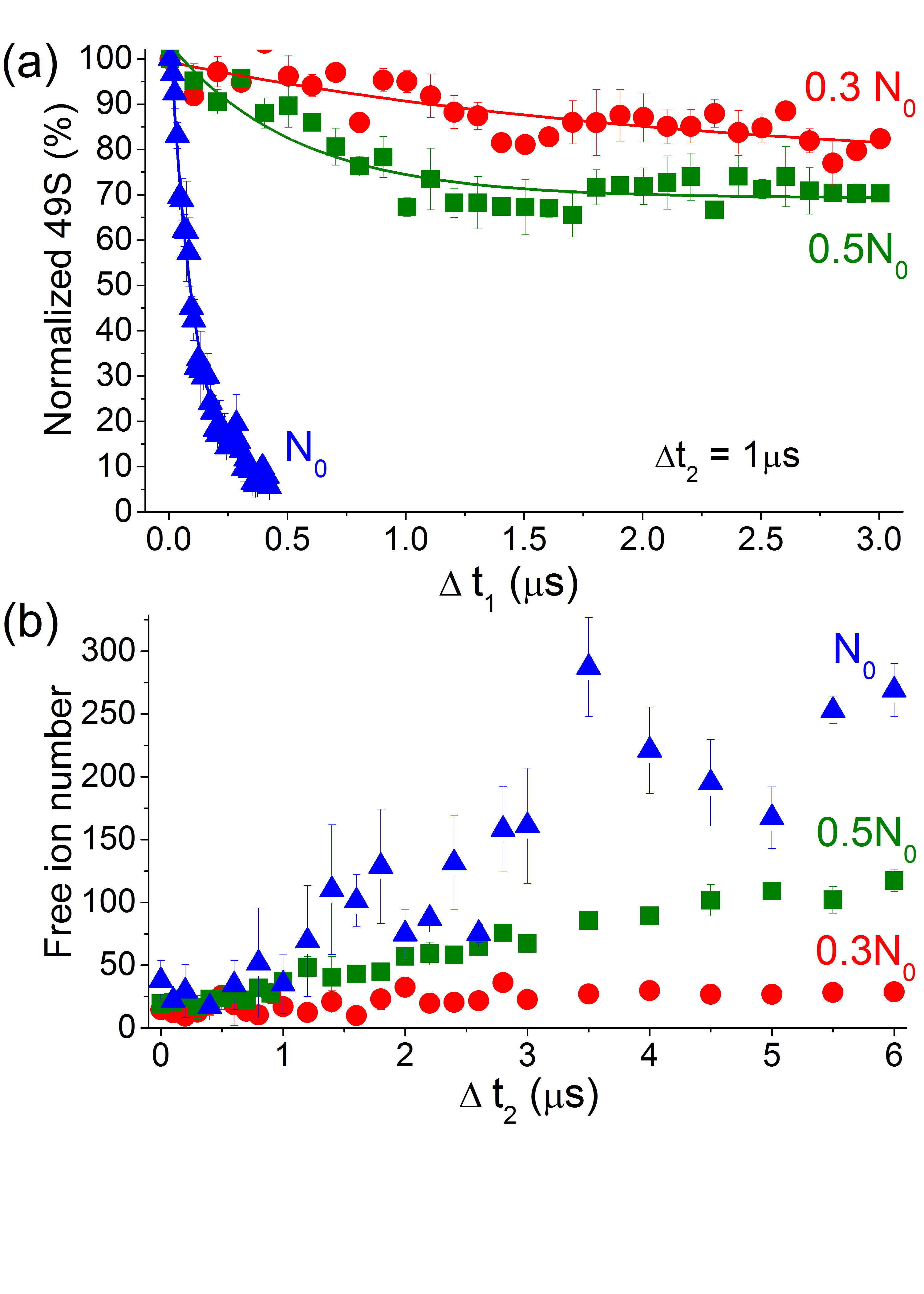}
%\vspace{-3em}
\caption{(Color online) (a) Measured normalized atom number of 49S$_{1/2}$ as a function of ramp time $\Delta t_{1}$ at different densities, $N_{0}$ (3$\times 10^{9}~$cm$^{-3}$), 0.5~$N_{0}$ and 0.3~$N_{0}$, of Rydberg atoms. The electric field is ramped  from 3.14 to 3.99~V/cm through one of the avoided crossings in Fig.~2~(a) for holding time $\Delta t_{2}$ = 1~$\mu$s. (b) Detected number of Penning-ionized atoms as a function of $\Delta t_{2}$ at $N_{0}$, 0.5~$N_{0}$ and 0.3~$N_{0}$ with ramp time $\Delta t_{1}$ = 80~ns for $N_{0}$ and $\Delta t_{1}$ = 500~ns for 0.5~$N_{0}$ and 0.3~$N_{0}$, respectively. In all cases, the $\Delta t_{1}$ are long enough to ensure primarily adiabatic passage.}
\end{figure}

As an additional consistency test of our interpretation, in Fig.~4~(b) we show the number of ions generated by Penning ionization as a function of $\Delta t_{2}$ for fixed $\Delta t_{1}$ and for different Rydberg atom densities. As the atom number and the atom density are lowered by about a factor of 3.3, the number of ions generated drops by about a factor of 12, while the curves remain linear (within our experimental precision). The drop factor and the linear time dependence of the ion signal are consistent with binary Penning-ionizing collisions generating the free ions.

\section{Landau-Zener dynamics}
\label{sec:lztheory}

The time scales on which the signals observed in Fig.~3~(b) and 4~(a) decay depend on both the passage behavior in the avoided crossing and the time required for atoms at point C in Fig.~2~(a) to become undetectable due to $m$-mixing. Atoms passing diabatically into point B in Fig.~2~(a) have wavefunctions very similar to those of atoms in point A. These do not efficiently mix and remain near-$100\%$ detectable for several $\mu$s, as shown in the inset of Fig.~3~(b). For the interpretation we have presented to hold, the ramp time $\Delta t_1$ for the passage to be primarily adiabatic must be less than the signal decay time seen in Fig.~3~(b) (which is 103~ns). In the following we show that this condition is met.

\begin{figure}[htbp]
\centering
\includegraphics[width=0.4\textwidth]{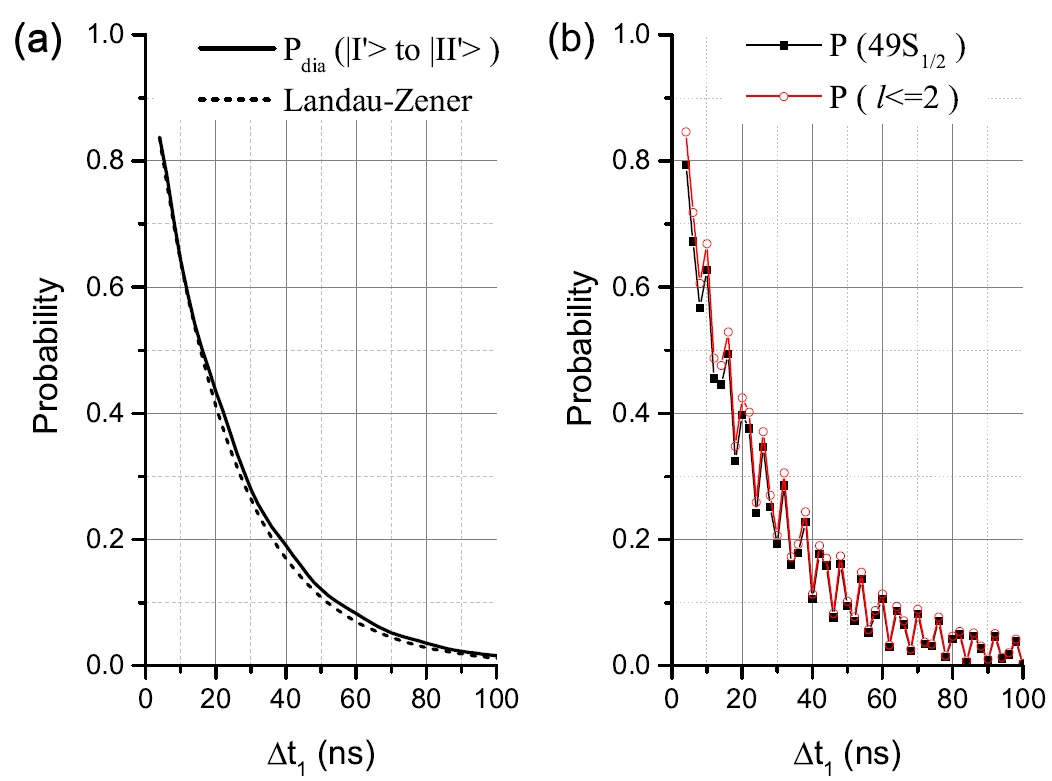}
\caption{(Color online) (a) Theoretical diabatic transition probabilities to tunnel from adiabatic state $\vert I' \rangle$ into
adiabatic state $\vert II' \rangle$ in Fig.~2~(a) as a function of the ramp time $\Delta t_1$, according to a simplified Landau-Zener model explained in the text (dashed) and a complete simulation (solid). (b) The squares show the probability of exiting the passage in the  49S$_{1/2}$-state and the circles the probability of exiting in any state with angular momentum \emph{l}$\leqslant 2$.}
\end{figure}

The avoided crossings under investigation involve three levels (see Fig.~2 (a)) and may therefore differ from a simple two-level Landau-Zener case. In Fig.~5~(a) we show the diabatic transition probability as a function of $\Delta t_{1}$, obtained by solving the time-dependent Schr{\"o}dinger equation for passage through the relevant adiabatic crossing in Fig.~2 (a). The exact diabatic transition probability is compared with an approximate diabatic transition probability derived from a simplified Landau-Zener model that only involves levels $\vert I' \rangle$ and $\vert II' \rangle$ in Fig.~2 (a). The close agreement between the exact and the approximate Landau-Zener transition probabilities shows that the intermediate adiabatic levels ($\vert III \rangle$ and $\vert III' \rangle$ in Fig.~2 (a)) are indeed only of minor importance. The close agreement also accords with the facts that the intermediate adiabatic levels do not noticeably repel from other levels in Fig.~2 (a), and that the calculated probabilities for the atoms to be in the intermediate adiabatic levels are $ < 1\%$.

According to the Landau-Zener model, the diabatic transition probability is given by
\begin{equation}
P_{\rm {dia}} =  \exp(-2 \pi \Gamma) \, {\rm ,where} \quad \Gamma = \frac{(\Delta E/2) ^2}{\vert (dF/dt) \cdot (d_{\rm I} - d_{\rm II})\vert} \label{eq:LZ}
\end{equation}
where $\Delta E$ is the gap size and $d_{\rm i}$ are the electric dipole moments of the coupled diabatic levels (all in atomic units). While the latter are not a priori known, the $d_{\rm i}$ are approximately given by the electric-dipole moments of the adiabatic states $\vert I' \rangle$ and $\vert II' \rangle$ at fields away from the crossing. Equation~(\ref{eq:LZ}) shows that the diabatic transition probability can be written as $P_{\rm {dia}} = \exp(-\Delta t_1 /\tau)$, with a characteristic time
\begin{equation}
\tau =  \frac{\Delta F \vert (d_{\rm I} - d_{\rm II})\vert}{2 \pi \, (\Delta E/2)^2} \label{eq:LZ2} \quad,
\end{equation}
where $\Delta F$ is the field range of the linear electric field ramp.
The values of $\tau$ depend on the gap size of the avoided crossing,
the differential energy slope of the crossing levels and the ramp parameters. The Landau-Zener curve in Fig.~5~(a) and Eq.~\ref{eq:LZ2} yield a characteristic time $\tau = 23~$ns, which is close to the
exponential-fit result for the complete simulation, $\tau = 24~$ns. These values are about a factor of four less than the 103~ns time scale observed for the signal decay in Fig.~3~(b) (which is for the highest-density condition we studied). This finding is consistent with our interpretation of the data. At higher Rydberg-atom densities, the observed signal decay time should approach the critical value $\tau = 23$~ns.

We expect minor deviations of the actual passage dynamics from the Landau-Zener case, as there are some deviations of our system from the assumptions made in the Landau-Zener model. Firstly, the diabatic states are a weak function of electric field, as evidenced by the fact that the electric dipole moments are slightly different before and after the crossings. Secondly, the weakly coupled third level will play a role in the case of very slow ramps.  Finally, we have also calculated the probabilities of exiting the avoided crossing in the state $49S_{1/2}$ or in any state with \emph{l}~$\leqslant 2$. These probabilities, displayed in Fig.~5~(b), exhibit modulations at a period of 6~ns. The modulations are due to quantum interference between the 49S$_{1/2}$ components present in both adiabatic levels populated after the passage.

\section{M-mixing}
\label{sec:mmixing}

For a duration of several microseconds, the Rydberg atoms primarily interact via the binary electric-dipole interaction,
\begin{equation} \label{eq:dipoleint}
    \hat{V}_{\rm{dd}}=\frac{{\hat{\textbf{p}}_1} \cdot {\hat{\textbf{p}}_2} - 3 ({\textbf n} \cdot {\hat{\textbf{p}}_1})({\textbf n} \cdot {\hat{\textbf{p}}_2})}{R^3} \quad .
\end{equation}
where the electric-dipole operators $\hat{\textbf{p}}_1$ and $\hat{\textbf{p}}_2$ act on the first and the second component of two-atom Rydberg states $\vert 1 \rangle \otimes \vert 2 \rangle$, respectively. There, the states $\vert 1 \rangle$ and  $\vert 2 \rangle$
are single-atom Rydberg states within the (non-zero) electric field $F$.
The unit vector ${\textbf n}$ points from the center of mass of the first to that of the second atom, and $R$ is the magnitude of the interatomic separation. The matrix elements are calculated as described in Ref.~\cite{AReinhard}.

To estimate the effect of $m$-mixing, we have integrated the time-dependent Schr\"odinger equation for atom pairs picked at initial nearest-neighbor separations for a density of $3 \times 10^9~$cm$^{-3}$, corresponding to our experimental conditions. For randomly positioned atom pairs, the initial atom separation follows a probability distribution
\begin{equation}
P(R)=\frac{3R^2}{w^3} \exp \left[- \left( \frac{R}{w} \right)^3 \right]
\end{equation}
where $w$ is the Wigner-Seitz radius. In the simulation the range of initial values of $R$ is restricted to $R>1.5~\mu$m, because lesser separations are unlikely due to the excitation blockade effect and Penning ionization. The initial angle $\theta$ of the internuclear separation vector relative to the electric field is chosen at random (with weighting $\propto \sin(\theta)$). The initial atom velocities are randomly selected from a Maxwell distribution (temperature $T$). The internuclear separation $R$ and its unit vector ${\textbf n}$ are a function of time, given by the randomly chosen initial values for positions and velocities. Due to the short interaction time, the effect of interatomic dipole-dipole forces on the (classical) internuclear trajectories are neglected. Since the internuclear separation vectors ${\bf {R}}$ depend on time, the matrix elements of the electric-dipole interaction operator in Eq.~\ref{eq:dipoleint} also depend on time.

\begin{figure}[htb]
%\vspace{-3em}
\centering
\includegraphics[width=0.4\textwidth]{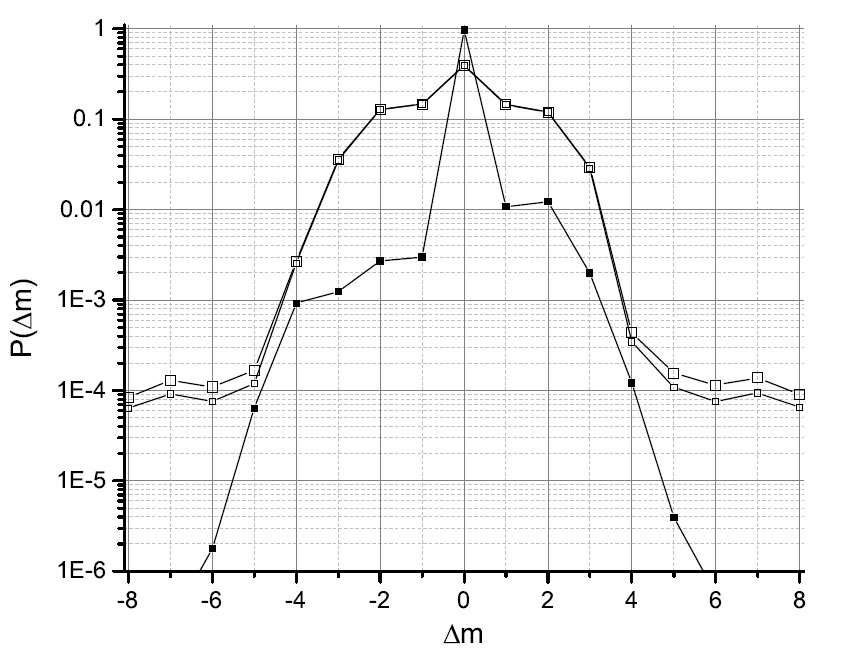}
\vspace{-0.0cm}
\caption{Probability for the change in single-atom $m$-quantum number due to binary electric-dipole interactions for atom pairs in the 49S$_{1/2}$-like state at point B in Fig.~2~(a) (solid squares; $T=200~\mu$K), and for atom pairs in the high-\emph{l} state at point C (open squares; large symbols: $T=200~\mu$K, small symbols: $T=1~\mu$K).
 All atoms in the simulation initially have $m_{0}$ = +1/2. For the high-\emph{l} state, the probability of $m$-change is much higher than for the 49S$_{1/2}$-like state.}
\end{figure}

The time-dependent Schr\"odinger equation is integrated over a wait time $\Delta t_2= 1~\mu$s, according to the experiment. The quantization axis is parallel to the electric field. The utilized internal-state space is restricted to two-body states for which both single-atom $m$-quantum numbers are within the range $m_0 - 8 \leqslant m \leqslant m_0 +8$, where $m_0$ is the initial-state $m$-quantum number (which is $ \pm 1/2$). Also, the effective quantum numbers of both atoms are within the range $43 \leqslant n_{\rm eff} \leqslant 47$. We run the simulation for two cases of the initial internal two-body state, namely $\vert B \rangle \otimes \vert B \rangle$ or $\vert C \rangle \otimes \vert C \rangle$, according to the points B and C in Fig.~2~(a). Finally, the two-body state space is restricted to states whose total energy differs by less than 40~MHz from the initial two-body state. This range of two-body states has been found large enough to yield convergent $m$-mixing probabilities.
Typically, there are several tens of thousands of two-body states in this range. From the final two-body state we extract the probabilities of the (single-atom) $m$ quantum numbers changing by amounts $\Delta m$. According to our basis restriction, $\vert \Delta m \vert \leqslant 8$. The simulation is repeated and the results are averaged over 1000 random choices of initial positions and velocities.

The selection rules of the electric-dipole interaction operator $\hat{V}_{\rm{dd}}$ in Eq.~\ref{eq:dipoleint} are, in first order, $\vert \Delta m \vert \leqslant 1$, for both of the atoms involved, and $\vert \Delta M \vert \leqslant 2$ for the sum of the two single-atom $m$-quantum numbers. The interaction time of 1~$\mu$s is long enough that $\hat{V}_{\rm{dd}}$ acts in higher order. As a result, large changes $\Delta m$ and $\Delta M$ are possible.

A typical simulation result is shown in Fig.~6. It is seen that the high-$\langle l \rangle$ elongated Stark states $m$-mix much more readily than the 49S$_{1/2}$-like ones. This may be expected because in the vicinity of point C in Fig.~2~(a) the density of higher-$\vert m \vert$ ``background'' states is larger than it is in the vicinity of point B. Also, the $\hat{V}_{\rm{dd}}$-couplings of the 49S$_{1/2}$-like atoms to other states are generally smaller, because at the small fields used in this work the $P$-character within the manifold of hydrogenic states is very small (due to the large quantum defect of Cs $P$-levels) and the admixture of high-$l$ character in the 49S$_{1/2}$-like levels also is small.  As a result of $m$-mixing, the high-$\langle l \rangle$ atoms selectively acquire higher field ionization thresholds.

The mixing percentages in Fig.~6 likely are underestimates because many-body effects beyond two atoms are not included; such effects have been found earlier to enhance near-resonant many-body mixing effects~\cite{Younge.2009}. Further, the effects of interatomic forces will lead to particle acceleration. We note the large permanent electric dipole moment of the high-$\langle l \rangle$ elongated Stark state, given by the large negative slope of the energy level at point C in Fig.~2~(a). The resultant strong interatomic permanent-electric-dipole forces may also enhance the $m$-mixing. Figure~6 shows that at higher temperatures there is slightly more $m$-mixing. This indicates that the mixing may accelerate once the sample heats up due to the interatomic forces.

\section{Conclusion}

Adiabatic/diabatic processes are ubiquitous in natural science; they are, for instance, manifested in atomic and molecular spectra, in collisions and in chemical reactions.
In this work, we have prepared high-density gases of strongly interacting Rydberg atoms based on avoided crossings formed by the 49S$_{1/2}$-state and $n=45$ hydrogenic states of cesium in an electric field. Electric-field ramps through the avoided crossings induce mixed adiabatic-diabatic passage behavior. In the adiabatic case, the atoms acquire large permanent electric dipole moments. The adiabatically transformed atoms are embedded in a background of other high-$\vert m\vert$ Stark states. These conditions are conducive to selective $m$-mixing of the adiabatically transformed atoms, making adiabatic passage experimentally detectable. Our measurements are in agreement with a model we have presented. In future work one may employ state transformation via adiabatic passage as a method to generate highly dipolar matter, against obstacles that may otherwise arise from inhomogeneous broadening, excitation blockade effects, and small rates for direct optical excitation of highly dipolar states. The oscillations seen in the calculations in Fig.~5~(b) are a general interferometric characteristic of systems involving two coupled quantum states; they may, in future work, enable measurements of dipole moments and other atomic properties.

\section*{Acknowledgements}

We thank Prof. J\"orn Manz for useful discussions. The work was supported by the 973 Program (Grant No. 2012CB921603), NSFC Project for Excellent Research Team (Grants No.61121064), NNSF of China (Grants No. 11274209, 61475090, 61378039 and 61378013) and NSF Grant No. PHY-1205559.

\end{document}